\documentstyle[12pt]{article}

\begin{document}
\bibliographystyle{unsrt}

\begin{flushright} 
UMD-PP-96-68\\
SMU-HEP-96-01\\
March, 1996

Astro-ph/9603049

\end{flushright}

\vspace{6mm}

\begin{center}

{\huge  Structures in the Mirror Universe}\\[6mm]

\vspace{6mm}

{\bf{R.N. Mohapatra\footnote{Work supported by the
 National Science Foundation Grant \#PHY-9119745 and a Distinguished
Faculty Research Award by the University of Maryland for the year 1995-96.}}}\\

{\it{ Department of Physics}}\\
{\it{University of Maryland}}\\
{\it{ College Park, MD 20742 }}\\

~~and~~\\

{\bf Vigdor L. Teplitz}\\
{\it Department of Physics}\\
{\it University of Maryland}\\
{\it College Park, MD 20742}\\
{\it and}\\
{\it Department of Physics\footnote{permanent address}}\\
{\it  Southern Methodist University}\\
{\it  Dallas, TX 28225  }

\end{center}

\vspace{4mm}

\begin{center}
{\Large \bf Abstract}
\end{center}
\vspace{1mm}

The idea of the universe with a mirror sector having all particles and forces
identical to those in the familiar sector
has been proposed in the context of neutrino physics as well as
superstring theories. Assuming that
all the quark and charged lepton masses 
in the mirror universe are scaled by a common factor,
$\zeta$, as is required in one interpretation of the neutrino data, 
we investigate domains of the parameter $\zeta$ where physical
conditions are favorable for cooling in the age of the universe that can
lead to the formation of compact structures given
the initial condition $\Omega_{B}\simeq \Omega_{\tilde{B}}$ ($\tilde{B}$
denoting the mirror baryon). In particular we ask whether there is a
region in $\zeta$-space for which primordial Jeans mass mirror clouds cannot
cool in the present age of the universe. We find that, for most of the area 
of interest in the parameter space, atomic hyperfine structure cooling is
effective in a time period short compared to the age of the universe but
long compared to the free fall time for globular cluster-sized objects
expected on the basis of simple Jeans length analysis.

\newpage

\noindent{\Large \bf I. Introduction:}

\vspace{4mm}

    The idea that our present universe may have a mirror partner
evolving with identical matter and force content has been invoked for
various reasons from time to time during the past thirty years.
The continuing interest in the phenomenology of such models was revived in the
late eighties by the observation that the superstring theories
(Green et al. 1986) lead
naturally to such a picture where
the known particles are accompanied by a duplicate set, with identical
properties, but in which the two sets have little or no interaction 
except for that of gravitation (see for example 
Khlopov et al 1991; Kolb, Turner and Seckel 1985; Hodges 1993 and references
therein).
Most recently this idea has emerged from attempts to understand the 
experimental observations relating to neutrino oscillations (Berezhiani and
Mohapatra 1995; Foot and Volkas 1995; Berezhiani et al. 1995)
such as the solar and atmospheric neutrino deficits (Winter 1995)
as well as indications
from the LSND (Athanassopoulos et al. 1995; Hill 1995)
data. The idea is particularly attractive in any
model that requires an ultralight sterile neutrino to fit
observations. Our focus in this paper will be on the class of models
advocated in Berezhiani and Mohapatra (1995) and 
Berezhiani, Mohapatra and Dolgov (1995)
where solar neutrino data require that the weak
scale of the mirror universe be higher than that of the standard model.
This model (the weak scale asymmetric mirror model, WSAM) 
has features of particular astrophysical interest: no 
mirror element, except hydrogen, is stable; the mirror proton 
mass is similar to that of the
normal proton; and the mirror electron masss is perhaps ten to 100 times
heavier than that of the normal electron, while the value of mirror 
electric charge is the same as for normal matter.  

	In Berezhiani et al. (1995), the 
possibility was raised that mirror matter might be detected by lensing or 
other gravitational phenomena, based on the mirror structures that might form,
such as supermassive mirror black holes, mirror machos, etc.  The purpose of
this paper is to study specifically the formation of mirror structures
in this model so as to permit more detailed investigation
of their implications for the model.

  The question of initial condensation and star formation 
of the primordial hydrogen cloud of the familiar universe has been a
subject of discussion among cosmologists for the obvious reason
of understanding the origin of first stars and galaxies (see for example, 
Peebles and Dicke 1968; Silk 1977; Palla et al. 1984; Tegmark et al. 1996).
Its study combines information from atomic physics and statistical physics 
as well as cosmology and chemistry. In our discussion,
we take over the relevant ideas with appropriately scaled
atomic parameters and combine them with the cosmology of the hidden mirror
world to get a qualitative understanding of structure formation in the 
(exclusively hydrogen) mirror sector. 
A significant difference is that, whereas 
 dust particles play a major role in understanding 
most of the detailed
structure in the familiar sector, the mirror sector 
for the range of interest of
the $\zeta\equiv m_{\tilde{e}}/ m_e$ parameter is devoid of any nuclei
heavier than the hydrogen and is thus free of dust particles. This to an
extent simplifies our discussion and enables us to draw the conclusions
that we do. 
Of particular interest to us is the nature and evolution of the 
low mass Population III stars for the visible universe.
Our calculations draw from the literature
on Population III stars, particularly the classic 1968 
paper of Peebles and Dicke.
	
The plan of the paper is as follows.  In Section II, we review the
necessary elements of the weak scale asymmetric mirror model(WSAM).  
In Section III, we
address cosmological structure issues, including mirror recombination,
primordial hydrogen molecule formation and the mirror Jeans mass.  It is not
surprising that, because of the similarity in proton mass, the mirror matter
Jeans mass at matter domination is about that of a globular cluster;
an important difference is that because of
the greater electron mass, mirror matter recombination precedes both that of 
normal matter and the point of matter domination over radiation.  In Section IV
we address the fate of Jeans mass "mirror globs."  We find that for most
regions of parameter space, the energy loss rate should be sufficient 
to support considerable condensation. We show that, as $\zeta$ increases,
molecular cooling becomes ineffective but cooling by the hyperfine structure
transition seems to take over sufficiently rapidly as to leave in all 
probability no interesting $\zeta$ region without a cooling mechanism.
In particular, we show that the rate of hyperfine cooling is small compared
to the free fall rate.
In Section V we discuss the fate of the globs, 
including possible scenarios within
the context of the results of Sections III and IV and possible observational
constraints. 

\vspace{4mm}

\noindent{\Large \bf II. Review of the Weak Scale Asymmetric Mirror Model} 

\vspace{3mm}

There exist several experimental indications of non-zero neutrino
masses and 
mixings: these include (i) the solar neutrino deficits 
observed now in four different
experiments, (ii) deficit of atmospheric neutrinos observed in IMB, Kamiokande
and Soudan II experiments and (iii) a preliminary 
indication of direct accelerator neutrino oscillation from the
LSND experiment (for a recent review, see Winter 1995). 
If all these data are confirmed by the planned
experiments such as SNO, Borexino and Super Kamiokande as well as the new
runs at LSND etc, they will have profound implications for the neutrino
mass matrices as well as for physics 
beyond the standard model. One simple way to
understand all these observations is to assume that there exists a fourth
ultralight neutrino species beyond the three already known (Caldwell and
Mohapatra 1993; Peltoniemi and Valle 1993) i.e. $\nu_e$, 
$\nu_{\mu}$ and $\nu_{\tau}$; in this picture, the solar neutrino puzzle is
understood via the oscillations between the $\nu_e$ and the extra neutrino
(denoted by $\nu^{\prime}_e$) both of these are assumed to have mass in the
milli-eV range. The atmospheric neutrino puzzle is solved via $\nu_{\mu}$-
$\nu_{\tau}$ oscillation. The present indications from LSND experiment
then imply that $m_{\nu_{\mu}}\simeq m_{\nu_{\tau}}\simeq $ few eV. Indeed
there exist analysis (Primack et al 1994)
of the existing data on structure in the 
universe which seem to require a hot dark matter component at the level of
20\% of the critical density which is fulfilled if the more massive of the
neutrinos above have a mass of $2.4$ eV each. 

The problem with such a picture is that since LEP data implies that only
three neutrino species can couple to the Z-boson, the fourth neutrino must
be singlet (or neutral) with respect to the standard model gauge group.
Since a standard model singlet fermion can in principle acquire an arbitrary
mass in the absence of any symmetry protecting its lightness, the puzzle
arises as to why such a neutrino (i.e. $\nu_e'$) is so light. It was
suggested (Berezhiani and Mohapatra 1995; Foot and Volkas 1995)
that if there is a mirror sector of the universe with identical
gauge structure and particle content to the standard model but decoupled
from the standard model particles except thru gravity, then the same 
symmetry (i.e. $B-L$) that keeps the ordinary neutrinos massless (or light) 
will have an analog in the mirror sector (to be called $B'-L'$) which
 will keep the $\nu_e'$ also massless or ultralight. It was suggested
(Berezhiani and Mohapatra 1995)
that the lightest neutrino masses arise from Planck scale
effects, in which case the masses of $\nu_e$ and $\nu_e'$ scale like the
square of the weak scales in their respective universes (i.e. $v_{wk}^2$
and $v_{wk}'^2$) whereas their mixing goes like $v'_{wk}/v_{wk}\equiv
\zeta$. From
the MSW solution to the solar neutrino puzzle, one can then conclude that
$\zeta\approx 30$. In this picture, the mirror analogs of $\nu_{\mu}$ and
$\nu_{\tau}$ will have masses in the range of few KeV's so that they can
constitute warm dark matter for the universe.

It furthermore turns out that since the value of $\Lambda_{QCD}$ in
the mirror sector is near that in the normal sector whereas the mirror
down quark, $d'$ is $\zeta$ times heavier than the $d$-quark, we expect
$m_{n'}-m_{n}\gg $binding energy of nuclei. As a result,
 the mirror neutron whether bound or free
is always unstable (Berezhiani, Dolgov and Mohapatra 1995)
 resulting in the startling conclusion that
in the mirror sector the only stable atom is the mirror hydrogen atom ($H'$).
The weak scale asymmetry model (WSAM) has also the major 
difference that the ionization energy of $H'$
is perhaps on the order of 400 eV (for $\zeta=30$).

The basic WSAM picture of the evolution of the mirror 
sector after the big bang
has been discussed in Berezhiani, Dolgov and Mohapatra (1995)
It was pointed out there that consistency
with standard big bang nucleosynthesis requires that there be asymmetric
reheating after inflation in the two universes with $T\simeq T'/2$. In fact,
in Berezhiani, Dolgov and Mohapatra (1995)
we constructed realistic models where the asymmetric
inflationary reheating is intimately connected with the asymmetry in the
electroweak scales. The rest of the history of the mirror sector can
be worked out until recombination in the mirror sector which takes place
when the temperature of the universe is around few tens of eV due to higher
ionization energy of the mirror hydrogen. The nature of structure in the
mirror sector depends on the detailed dynamics of the mirror
sector and is the problem we attempt to tackle in this paper.  

\vspace{4mm}

\noindent{\Large \bf III. The Early Years}

\vspace{3mm}

In attempting to study possible structure formation in the mirror
universe in WSAM, we follow a scenario similar to the familiar sector
(Kolb and Turner 1990; Weinberg 1972; Peebles 1993 and references therein).
For $\zeta$ mildly greater than one ($\zeta\geq 3$), mirror
recombination takes place before matter dominance.
After the mirror recombination
(at which point the mirror baryon component dominates the expression
for Jeans length), the mirror universe contains
a neutral hydrogen cloud with a small fraction of ionized hydrogen 
and electrons. When the universe later 
becomes matter dominated, the density fluctuations grow 
with time until the expansion of the high density regions
stop and violent relaxation takes place leading to isolated "globs".
The future evolution of the globs is dictated 
by the rates for physical and chemical
reactions that determine the extent of energy loss by the globs. If the
energy loss rate is significant, then condensation to 
jupiter type objects of low mass or supermassive black holes 
can proceed; otherwise the gas cloud remains diffuse making detection
difficult. The primary difference between the mirror and the familiar
sector will be the scale factor
$\zeta$ which will make the reaction rates in the mirror universe
different.
 
	We compute: (1) amount of ionization at mirror matter
recombination; (2) density of primordial mirror hydrogen molecules; (3) Jeans
mass and length of a mirror matter "glob" at matter domination; and (4) glob
parameters after violent relaxation.  We consider, for definiteness in most
places, that the
universe has the critical density, with normal and mirror baryons each
providing 0.05 and mirror neutrinos, the model's warm dark matter, the
remainder.  We take the Hubble constant to be 50 km/s-Mpc.  We take the ratio
of the mirror proton mass to that of the normal proton to be $\alpha=1.5$ 
and the ratio
of the mirror temperature to the normal temperature to be $\beta=1/2$.  
We sometimes
keep the ratio of the mirror electron mass to that of the normal electron
a free parameter $\zeta$; where a definite value is useful, we choose $\zeta=
 30$.

	In computing recombination in the mirror model, we
follow the treatment of Kolb and Turner (1990) for normal matter, 
and reproduce it in order to compare the mirror matter calculation with it.  
We consider the reaction:
\begin{eqnarray}
e^-+p\rightarrow H~+~\gamma
\end{eqnarray}
We define the variable $x=B/(2T)$, where $B$ is 13.6 eV.  
The reaction rate for (1) is then given by:
\begin{eqnarray}
\langle \sigma v\rangle = A_1\left( x^{1/2}/ m^2_e\right)
\end{eqnarray}
with $A_1={{4\pi^2\alpha^2}\over{\sqrt{3}}}\equiv 1.46\times 10^{-14}~ cm^3 
s^{-1} MeV^2$.
We have for the equilibrium ionization fraction,
\begin{eqnarray}
X_{eq}= 0.51 \eta^{-1/2}\left({{m_e}\over{T}}\right)^{3/4} e^{-x}\\
= A_2 x^{3/4} e^{-x}/(\Omega^{1/2}_B h)
\end{eqnarray}
with $\eta=(\Omega_Bh^2) 2.68\times 10^{-8}$ and $A_2=1.41\times 10^7$.
We have for the density of electrons,
\begin{eqnarray}
n_e~=~\eta n_{\gamma}X_e~=~ A_3 x^{-9/4}m^3 e^{-x} (\Omega_B h^2)^{1/2}
\end{eqnarray}
with $A_3=2.8\times 10^{16}~cm^{-3} MeV^{-3}$.
Finally, the time of recombination is
\begin{eqnarray}
t_R={{2}\over{3}} H^{-1}_0(T_0/T)^{3/2}~=~ A_4 x^{3/2}/(h m^{3/2})
\end{eqnarray}
with $A_4= 1.55\times 10^{10}~MeV^{3/2}s$.

Setting $\langle \sigma v\rangle n_e t_R=1$ to find the time at which 
freezeout occurs gives the condition,
\begin{eqnarray}
A_1A_3A_4 m^{-1/2} x^{-1/4} e^{-x}\Omega^{1/2}_B = 1
\end{eqnarray}
Using this one finds, $x^{-1/4} e^{-x}= 1.56\times 10^{-13}\left({{m}\over
{\Omega_B}}\right)^{1/2}$. This gives for the fraction $f_e$ of free electrons
\begin{eqnarray}
f_e=X_{eq}= {{A_2m^{1/2}x}\over[A_1A_3A_4\Omega_Bh]}=
6\times 10^{-5} {{m}\over{\Omega_Bh}}\left({{x}\over{27.5}}\right)
\end{eqnarray}
This is in rough agreement with Kolb and Turner (1990).

	In the WSAM model, we must modify
 reaction rates appropriately. Denoting all parameters of the mirror
sector by a tilde and defining $\alpha= m_{\tilde{p}}/ m_p$ and 
$\beta=\tilde{T}/ T$ and $\zeta={{m_{\tilde{e}}}\over{m_e}}$ as before, 
we have for the reaction rate for the analog of reaction (1)
\begin{eqnarray}
\langle \sigma v \rangle = A_1 \tilde{x}^{1/2}/\tilde{m}^2_e
\end{eqnarray}
where we have assumed that the cross section for (1) scales with the square of 
the Bohr radius and $\tilde{x}= \tilde{B}/2\tilde{t}$.  
The analogues to Equations (3-6) are 
\begin{eqnarray}
\tilde{X}_{eq} = A_2 \alpha^{1/2} \tilde{x}^{3/4} 
e^{-\tilde{x}}/(\Omega^{1/2}_{\tilde{B}}h)
\end{eqnarray}

\begin{eqnarray}
{\tilde{n}}_e = A_3 \tilde{x}^{-9/4}\tilde{m}^3 e^{-\tilde{x}}
(\Omega_{\tilde{B}}^{1/2}h)/\alpha^{1/2}
\end{eqnarray}

\begin{eqnarray}
t_{\tilde{R}}= A_5 h^{-1} \left( {{\tilde{x}\beta}\over{\tilde{m}}}
\right)^2
\end{eqnarray}
where $A_5= 9.6\times 10^9 MeV^2s$. 
Setting $1=\langle \tilde{\sigma}\tilde{v}\rangle \tilde{n}_e t_{\tilde{R}}$
gives

\begin{eqnarray}
\tilde{x}^{1/4} e^{-{\tilde{x}}}=\left( {{\tilde{m}\alpha^{1/2}}
\over{\Omega^{1/2}_{\tilde{B}}[A_1 A_3 A_5h \beta^2]}}\right)
\end{eqnarray}
so that we have

\begin{eqnarray}
\tilde{X}_{\tilde{R}}/ X_R =(A_4/A_5)\left({{m^{1/2}\tilde{x}^{1/2}}
\over{x}}\right) \left(\zeta \alpha/\beta^2h\right)\simeq 2.5 \zeta
\end{eqnarray}

	Numerically, Eq(14) corresponds, 
with our standard parameters ($\zeta=30$, etc), to about  
10\% unrecombined mirror electrons.

	We now use this result to compute primordial formation of mirror matter
molecules (Hydrogen).
Primordial molecule formation has been addressed in detail 
by Lepp and Shull (1984) and in summary by Peebles (1993)
 $ H_2$ is formed in the early universe,
before galaxy formation, principally in two catalytic processes
\begin{eqnarray}
e^-+H\rightarrow H^-+\gamma
\end{eqnarray}
followed by				
\begin{eqnarray}
H+H^-\rightarrow H_2+e^-
\end{eqnarray}
and
\begin{eqnarray}
p+H\rightarrow H^+_2+\gamma
\end{eqnarray}
followed by
\begin{eqnarray}
H+H^+_2\rightarrow H_2 + p
\end{eqnarray}
In each case, the first reaction is slow and the second fast.  In each case,
the amount of $H_2$ formed is determined by the temperature at which the CBR
can no longer dissociate the ion ($H^-$ or $H^+_2$ ).
Lepp and Shull give the two
temperatures as $2.75Kz_{eff}$ with $z_{eff}$=64 and 190, for $H^-$ and 
$H^+_2$, respectively; the
corresponding binding energies are 0.75 eV and 2.65 eV.  The amount of $H_2$
produced, in either case, is then
\begin{eqnarray}
f_2~=~t(z_{eff}){{df_2}\over{dt}}~=~\langle \sigma v\rangle (z_{eff})
n_H(z_{eff})f_e t(z_{eff})
\end{eqnarray}
where t is the age of the universe at $z_{eff}$. $f_2$ in 
Eq (19) rises with $z_{eff}$
since the $z^3$ behavior of $n_H$ outweighs the $z^{-3/2}$
behavior of $t(z_{eff})$.  For
familiar $H_2$ , Lepp and Shull show that, although 
$z_{eff}$ is larger for Eq (19), the
contribution of Eq (15) is about three times as big owing (mostly) to the
smaller velocity of the proton in Eq (17).  For mirror matter, however, the
velocity of the electron falls like $\zeta^{-1/2}$ , 
so we expect Eq (19) to be a good
approximation and dominant for  $\zeta \geq 9$. $\langle \sigma v \rangle$
for familiar matter, for Eq (19) is a
constant ($1.4\times 10^{-18} cm^3s^{-1}$ is used by Lepp and Shull (1984))
 at low energy, so a conservative estimate for mirror matter is simply
to scale it as $\zeta^{-2}$ .
$z_{eff}$  scales as $\zeta/\beta$ since
$\tilde{T}=\beta T$, so we take $z_{eff}= 380\zeta$        .
Thus, for mirror matter, Eq (17) becomes
\begin{eqnarray}
\tilde{f}_2=\tilde{f}_e \langle \tilde{\sigma} \tilde{v} \rangle n_{\tilde{H}}
(z_{eff}) t(z_{eff})
\end{eqnarray}						
with $\tilde{f}_e= 4\times 
10^{-3} \zeta$ and $t(z)= 4\times 10^{17}/z^{3/2}$,
 giving
\begin{eqnarray}
\tilde{f}_2~=~1.2\times 10^{-6}\zeta^{1/2}
\end{eqnarray}
This is about 60 times the amount for familiar matter,
with the same $\Omega$, for this process.  It would be of interest 
to compare this
rate with results for muonic molecules; however the reactions of interest for
muon catalysis appear to be only those involving atoms, rather than ions, since
densities are high (Hughes and Wu 1977).  
On the other hand, the precise numerical result in
Eq (19) will not be needed below.  In Section IV, we will note that the
most important
factor in molecular cooling, as a function of zeta, is that, for given
zeta, the temperature be high enough to permit exciting molecules out of the
rotational ground state.

	Third, we turn to the Jeans mass.  We consider, as noted, a three
component system - massive mirror neutrinos dominating the mass density,
together with both normal and mirror baryons.  The "baseline value" for matter
domination is z=5800; and the eigenvalue Jeans equation is
(Kolb and Turner, 1990)

\begin{eqnarray}
\delta_i v^2_i k^2 - 4\pi G_N \Sigma_j \rho_j \delta_j~=~0
\end{eqnarray}
where $\delta_i$ are the density fluctuations in the various components,
($i={\tilde{\nu}}, \tilde{B},B$)
$\rho_i$ are the mean densities, $v_i$ are the sound velocities in the
different media and $k$ is the wave number for the density fluctuation.
The various velocities are given by $v_{\tilde{\nu}}=\sqrt{3}\tilde{T}/
m_{\tilde{\nu}}$, $v_{\tilde{B}}=\sqrt{3}\tilde{T}/(m_{\tilde{B}}
\tilde{T}_{rec})^{1/2}$, $v^s_{B}=1/\sqrt{3}$

The solution to Eq (22) is

\begin{eqnarray}
\delta_i= A/{v_i}^2;~~ k^2= 4\pi G\Sigma_j\rho_j/v^2_j\equiv \left({{2\pi}
\over{\lambda}}\right)^2
\end{eqnarray}
	We note that the other eigenvalues of Eq. 
(22) are all zero, for any number of components, and that the amplitudes in the
solution of Eq. (23) are driven by the inverse of the sound speeds; the
densities do not enter.  The evolution of galaxies presumably proceeds at
leisure as in warm mixed dark matter models; see, for example, Colombi,
Dodelson and Widrow (1996) and references therein for recent detailed
discussion.
It is clear that, as the coolest component by far, the mirror baryons dominate
the solution.  We find a Jeans length and mass of

\begin{eqnarray}
\lambda_J\simeq 7\times 10^{18}(z_6/z_{MD})^{1/2}\tilde{\Omega}^{-1/2}_J~cm
\end{eqnarray}

\begin{eqnarray}
M_J=2.9\times 10^{38}\tilde{\Omega}^{-1/2}_J(z_{MD}/z_6)^{3/2}
\end{eqnarray}
where $z_6= 5.6\times 10^3$ and $\tilde{\Omega}_J=\Omega_{\tilde{B}}/0.05$
The mass in Eq. (25) is solely that of the mirror baryons within the Jeans
length.  The basic result is, not surprisingly, on the order of a globular
cluster. In Table 1, we collect some of the features of the early thermal
history.

	Finally, we need to make assumptions about the rate of growth of the
inhomogeneity.  We adopt a simple parametrization.  We assume that the mirror
hydrogen globular cluster size glob continues to expand with the rest of the
universe from the time at z=5800 of initiation of matter domination until 

\begin{eqnarray}
z_{stop}\equiv 5800 z_M
\end{eqnarray}
At that point "violent relaxation" begins.  It leads to glob of half the
original($z_{stop}$) size with temperature 
determined by the virial theorem.  The Mass,
radius, Temperature, and mirror particle density in the glob, and the age of
the universe are given, in terms of the parameter $z_M$, in Table 2.
There will, of course, be a distribution in mass of the structures; we
restrict this first look at the structure question to just those of
approximately the Jeans mass.

	The prediction of the mirror matter model is then 
formation of structures roughly on the order of those shown in Table 2.  There
is a strong dependence on the choice of the parameter $z_M$ 
in the table.  We, of
course, do not know the rate at which inhomogeneities in mirror matter will
grow, but we do know that they should become nonlinear
 earlier than those of
normal matter, since mirror recombination is earlier (and indeed before
the time of matter domination) and since $\delta_i$
  in Eq (23) is
significantly larger because of the smaller sound velocity.  In the estimates
below, we shall consider 0.1 and 0.01 values that can't be ruled out for $z_M$,
but not necessarily maximum and minimum values.

\vspace{4mm}
\noindent{\Large \bf IV. Energy Loss}
\vspace{3mm}

	In this section we estimate rates at which the mirror matter globs can
lose energy by different processes so as to determine how compact present day
objects might be in order to estimate their susceptibility to gravitaional
detection.  We consider three processes: (1) free-free radiation -- as modified
by continuing recombination; (2) the atomic mirror hydrogen hyperfine structure
line (or 7 mm line); and (3) the ortho-para mirror molecular rotational 
transition line.  The basic tension in cooling is that the temperature scale is
set by the time of matter domination and so is somewhat similar to that for 
normal matter; but the processes are all strongly dependent on the mirror 
electron mass which is greater by 10 to 100 than for normal matter.  There is
also strong dependence on the parameter $z_M$  
which, as seen in table 2 enters in
most quantities of interest.  We make estimates mostly on the
basis of constant (average) densities, but consider where necessary
modifications for an isothermal sphere. Note that, since in the mirror
sector there is no stellar burning, there is in all likelihood no 
reionization and we do not consider such a possibility.

	We begin with the question of glob recombination which will determine
the availability of mirror electrons for free-free radiation.  Using Eq (9),
with T as in Table 2, we can write for the time rate of change of the mirror
electron fraction in the glob

\begin{eqnarray}
{{df_e}\over{dt}}= - n_H f^2_e\langle \sigma v\rangle =-f^2_e/t_q
\end{eqnarray}
where $t_q\simeq 10^{10}z^{-5/2}_M~sec$. The solution is

\begin{eqnarray}
f_e= f_{e,0}/[ 1+ t f_{e,0}/t_q]
\end{eqnarray}
We thus see that free electrons will be reduced by at least a factor of ten in
time $t_0$ , the current age of the universe, so long as $z_M$ 
is greater than 0.01, and 
considerably more for higher values.  We note that the gammas from glob
recombination do not constitute a thermal energy  mechanism for the glob
since the Thomson cross section is too small for them to heat the remaining
electrons, so that they either remove binding energy from the glob in the case
of recombination to excited states or reionize mirror atoms in the case of
gammas from recombination to the 1S state too far from the glob surface.

	The rate at which the electron gas can absorb energy T from an atom is
given by
\begin{eqnarray}
t^{-1}_1\simeq f_{\tilde{e}} n_{\tilde{H}} 
\tilde{\sigma}v_{\tilde{e}}\left( {{m_{\tilde{e}}}\over{m_{\tilde{H}}}}
\right)^2\simeq (\zeta/10)^{-1/2} 10^{-10}f_{\tilde{e}} z^{7/2}_M~sec^{-1}
\end{eqnarray}
where we have estimated the cross section as being just the mirror bohr area,
Eq. (27) assumes that there is no problem with radiating the energy; it just
computes the time needed to transfer the energy from an atom to the free
electrons (in fact the bremsstrahlung rate will fall as $\zeta$ increases).
  The result is that, although fewer scatterings are needed as zeta
increases, the cross section falls so that the time required is 
nearly independent of zeta.
Note that the factor of $(m_{\tilde{e}}/m_{\tilde{H}})^2$  
represents the random walk in energy space needed.  The
actual rate of energy loss will be still less because of opacity, but Eq (29)
shows, taking into account the decrease in $f_{\tilde{e}}$
 of Eq (28) as $z_M$ increases from $0.01$, that significant
condensation via free-free radiation must take more than the age of the 
universe.  Note that, since we have $\sigma \approx \zeta^{-2}$ , 
$f_{\tilde{e}}\approx \zeta$, $v\approx \zeta^{-1/2}$ in Eq.(29),
this conclusion is independent of $\zeta$. This argument also indicates
that Compton cooling  (scattering of electrons off CBR photons)(see Silk,1976)
cannot cool either in the parameter range under consideration. 

	Next, we turn to a particularly interesting possibility, open to mirror
hydrogen because of the increased energy per transition and the increased
transition rate that come with the greater electron mass-- the atomic hyperfine
transition.  The time needed to radiate an atom's total kinetic energy in units
of the hyperfine transition energy is given by
\begin{eqnarray}
t_{2, R}\simeq \left({{3/2 T}\over{\epsilon^N_{HF}\zeta^2}}\right)({{t^N_{HF}}
\over{\zeta}})\simeq 10^{21} z_M/\zeta^3~sec
\end{eqnarray}					
where the factor in the first set of parentheses is the 
number of transitions each atom needs to
make, while $\epsilon^N_{HF}$ and $t^N_{HF}$ 
are the energy and inverse radiation rate for the case
of normal matter. Here we have used the fact that the hyperfine splitting
goes like $m^2_{\tilde{e}}/m_{\tilde{N}}$.
 We can derive a second characteristic time in addition
to that of Eq.(30) which is the total time required for the atoms to make the
transitions that radiate away their kinetic energy. The second characteristic
time is that required to excite the atoms to the higher energy state so that
they can do the radiating. The reason for treating the two times separately
is that each must be less than the age of the universe in order for the glob
to cool by this process. The total excitation time is given by
\begin{eqnarray}
t_{2,E}\simeq \left({{3/2T}\over{\epsilon^N_{HF}\zeta^2}}\right)(n_{\tilde{H}}
\sigma_{\tilde{H}\tilde{H}}v_{\tilde{H}})^{-1}\simeq 10^{10} z^{-5/2}_M sec
\end{eqnarray}
In Eq. (31), we have taken the cross-section to be six Bohr areas (Allison and Delgarno 1969).
We see from Eq (30)
that energy loss from the glob, in less than the age of the universe, tends to 
be possible for small values of $z_M$  and
large values of $\zeta$ that is, when the 
kinetic energy to be dissipated is small
and the energy dissipated per transition is large.  We note, however, that our
calculation is a simple-minded one; we have not considered such possible
additional effects as collisional de-excitation. We note that
for $\zeta > 1000$, the
energy difference between the two levels would increase to the eV range
thereby making it difficult to excite the atom to the higher level (i.e.
making $t_{2,E}$ larger), thus disfavoring this as a cooling mechanism.

	Finally, we turn to cooling by molecular radiation.  This is the chief
source of cooling in the Population III globular cluster work, for example
of Peebles and Dicke (1968) and Tegmark et al. (1996).
 In the present case its effectiveness is 
enhanced (for $\zeta >1$)
by the increased energy carried away in each transition, but is decreased by
the difficulty of exciting rotational states, since even the lowest-energy
excited state soon becomes more than the 
temperature that corresponds to $z_M   =0.1$. 
We can write for the time needed to remove energy 3T/2 from every atom

\begin{eqnarray}
t^{-1}_3\simeq \Lambda_2 \langle \sigma v\rangle f_2 \left({{\Delta E}\over
{3/2 T}}\right)\simeq 7\times 10^{-11}\zeta^{1/2}z^{5/2}_M \Lambda_2 ~sec^{-1}
\end{eqnarray}
where we use again the bohr area for $\sigma$, use $f_2$ 
as computed in Section II, and
have $ 0.015 eV\zeta^2$ for $\Delta E$ using the experimental value for
the ortho-para transition for $\zeta=1$. The lowest excitation energy is
three times the size of the "average" rotational energy ($(m_e/m_H) E_{el}$);
characteristic energy levels are given in Table 3.  
The factor $\Lambda_2$  takes account of the fraction of
collisions that are sufficiently energetic to excite the J=0 to J=1, para to
ortho, hydrogen molecule transition, and is given by
\begin{eqnarray}
\Lambda_2={{\int{d^3v_1 d^3v_2 e^{-{(2mv^2_1+mv^2_2)/2T}}\theta(m/3(v_1-v_2)^2-
\Delta E)}}\over{\int{d^3v_1d^3v_2e^{-(2mv^2_1+mv^2_2)/2T}}}}
\end{eqnarray}

Eq. (33) is written for collisions of H atoms
with $H_2$
molecules (hence the factor of 2 in the exponent);
however $\Lambda_2$ is independent of the masses of the particles and
depends only on $\Delta E/T$.
Evaluating Eq.(33) numerically, we find, approximately, that the time in 
Eq.(32) is greater than the age of the universe for the curve $t_3=t_0$ in
Fig.1.
That is, the glob can lose energy by molecular 
radiation for  larger values of  $z_M$
and smaller values of $\zeta$  , 
since in this case it is the ability to excite that
is important. The range of $\zeta$ for which molecular radiation can be 
effective will be further limited by the decreasing (with $\zeta$) 
cross-section for collision induced radiation
but it will not be necessary to compute this additional effect
nor do we compute the extra time needed to excite to $J=2$ state so
that radiative quadrupole radiation can occur without collision.

	We now consider the question of opacity.  For the case of the
rotational line, following Lang (1980) and Rohlfs (1986), 
we have for the opacity.

\begin{eqnarray}
\kappa_\nu= {{c^2}\over{8\pi \nu^2}}f_2 n_H A[1- e^{-\nu/T}]\phi(\nu)
\simeq 3\times 10^{-18}  n_H/z^{1/2}_M
\end{eqnarray}
where $A$ and $\phi$  are given by
\begin{eqnarray}
A={{32\pi^4}\over{3c^2}}\alpha\nu^3 a^2_B/\zeta^2
\end{eqnarray}
and
\begin{eqnarray}
\phi(\nu)={{c}\over{2\nu}}\left({{M}\over{2Tln2}}\right)^{1/2}\simeq 10^4/z^{1/2}\nu
\end{eqnarray}
From Eq (34) one can check that 90\% of the mass is within an optical depth of
300 from the surface so that total radiation time does not become a limiting
factor in energy loss.  One can also see from Eq (35) that, for the case of the
magnetic dipole hyperfine structure line the opacity will be reduced by at
least a factor of $10^{5}$   so that the optical depth will be less than one.

We have found that: (i) free-free radiation cannot give enough energy loss
to permit collapse for any $\zeta$, for reasonable $z_M$ ; 
(ii) hyperfine structure radiation cannot give such 
energy loss if $\zeta$ is too low, for reasonable $z_M$ 
  because the rate at which
energy is radiated is too small and it cannot give such energy loss if zeta is
too high (over 1000), for 
reasonable $z_M$, because the rate at which the upper hyperfine
state is excited becomes too small; and, (iii) finally, 
molecular radiation cannot
give such energy loss if zeta is too high (over 5), 
because the ortho to para molecular
transition cannot be excited with the available thermal kinetic energy.  These
results are summarized in Figure 1 where we show the three curves
\begin{eqnarray}
t_{2,R}\equiv 10^{21}z_M/\zeta^3 = t_0\equiv 4\times 10^{17} sec
\end{eqnarray}

\begin{eqnarray}
t_{2,E}= 10^{10} z^{-5/2}_M/ \Lambda_1 =t_0
\end{eqnarray}
where $\Lambda_1$ is the analogue of Eq.(33) for $\tilde{H}\tilde{H}$
scattering. Finally, the condition $t_3= t_0$ leads to
\begin{eqnarray}
z^{5/2}_M \Lambda_2 = 1.2\times 10^{-8}
\end{eqnarray}

	In Fig 1, the curve left-most at the bottom ($t_3=t_0$) is that 
above which the glob can cool
by means of the ortho-para $H_2$
transition; below it, the temperature of the
glob after violent relaxation is too low to excite the transition.  The second
curve ($t_{2,R}=t_0$)
is that below which the glob can radiate away its energy by means of the
hyperfine transition; above it, too many transitions are needed to radiate away
the thermal energy in time $t_0$.  The rightmost curve is that above which
the glob can excite atoms to the upper hyperfine state a sufficient number of
times to radiate away the thermal energy; below it the density and excitation
rate are too low.  In the shaded area between the first two curves and
 under the third curve
cooling cannot take place. Our result is that cooling is likely possible
for most of the region $\zeta \leq 400$ and $z_M\geq 0.01$  with the
lower bound on $log z_M$ rising approximately like $1.6 log \zeta$.

\vspace{4mm}
\noindent{\Large \bf V. Where Are They Now?}
\vspace{3mm}

	Given the above results, we can now ask the likely fates of mirror
globs and the possibilities for detecting signs of their existence.  There are
three broad possibilities: (i) Puffy Globs; this would be the case if we are in
a region of the $z_M-\zeta$ 
plane in which the globs cannot lose energy within $t_0$.
(ii) Cluster Globs; this would be the case for a region of the $z_M-\zeta$
 plane in
which the globs can lose energy sufficiently rapidly, and there is
fragmentation to low mass bodies, which would essentially be white dwarfs
or "small" black holes, since no nuclear burning is possible.  (iii) Black 
Globs; this would be the case for a region of the $z_M-\zeta$     
plane in which there is 
energy loss and no fragmentation, and hence collapse to a supermassive black 
hole.

	Case (i) will obtain for the shaded regions of Figure 1, providing
there are not important energy loss mechanisms we have neglected.  Case (ii),
in principle, could obtain for portions of the $z_M-\zeta$ 
plane in which the energy 
loss is through the hyperfine line.  This is because the glob will be optically
thin to the hyperfine line and hence, if the energy loss were rapid enough, 
particle velocities will decrease and regions of smaller Jeans length will 
appear.  See, for example, the discussion of Bohm-Vitense (1989), 
or the original
paper of Hayashi (1966).  However, the rate of energy loss is never rapid enough
from the hyperfine line.  That is, we can take the total time for energy loss
by adding Eqs. (37) and (38), divide by the free fall time, $(G\rho)^{-1/2}$,
and minimize the ratio with respect to $\zeta$.  We find

\begin{eqnarray}
(t_{2R}~+t_{2E})/t_{ff}\geq 10
\end{eqnarray}

Were the left hand side of Eq. 40 much less than one, we could expect
fragmentation through decreasing Jeans mass leading to subregion collapse.
  Case (ii) will also obtain if the mechanism of Peebles 
and Dicke (1968) for sequential central star formation in globular clusters
(followed by glob reheating and radiation from central objects formed) is 
applicable, or if other energy loss mechanisms turn out to be important.
The mechanism of Peebles and Dicke might maintain case (i) for the current age
of the universe.
Knowledge of star formation, for ordinary matter,
especially formation of Population III stars, is not sufficiently advanced, as
far as we can determine, to draw any confident conclusions.  Case (iii) will,
of course, obtain where there is energy loss without fragmentation to case
(ii).

	Given these three cases, we can look a little further into the fate of
the globs by asking about glob-glob scattering under the assumption that the
globs are bound to galaxies.  We take a simple model of our galaxy as a sphere
of radius 70 kpc, with a virial velocity on the order of $v_g= 10^{-3}c$
 The number of collisions a glob is likely to make is then given by
\begin{eqnarray}
N_c= ( N_g/V_g) \sigma_g v_g t_0\approx 10^{-3}z^{-2}_M
\end{eqnarray}
where we have assumed $N_g=10^6$ globs in 
the galaxy, and taken as the cross section
the area corresponding to the glob radius in Table 2.  We have neglected
gravitational focusing.  We can, in addition, ask about the angle of
scattering.  We have (see, for example, Binney and Tremaine 1989)
\begin{eqnarray}
tan \theta \approx \Delta v/v ~=~2GM/(b v^2)\approx 0.1 z_M
\end{eqnarray}

The results of Eqs. (39) and (40) are that we are near a borderline; we might
have a lot of scattering, including some 
large angle scattering that could liberate
the glob from the galaxy.  It would appear unlikely that we would have less
than on the order of a percent of the globs liberated.  Thus,
one might consider
searches for shadow matter globs both inside the galaxy and outside.  Searches
inside benefit from the possible existence of a large number of globs close at
hand.  Searches outside benefit from the large ratio of globs to galaxies, even
with only 1\% liberation.

	We note that we would expect, in case (ii) above that some percentage
of the cluster elements would be liberated in glob-glob scattering, but we
would consider it unlikely that a large fraction of mirror matter goes into
making MACHOs.  We should also note that cases (i) - (iii) above are not
necessarily mutually exclusive: while there will only be one value of zeta, we
expect some range of values of $z_M$ for globs.  If near a border, these could
include more than one case in the above taxonomy. Finally glob-glob mergers
could play an important role in determining some mirror sector structures
and should be considered in more detail.

	Finally, we turn to the question of potential observational
constraints.  The one that comes most immediately to mind is that of lensing,
galactic and extragalactic; detailed analysis is needed and possible based on
the above results, but is beyond the scope of the present paper.  There are,
in addition, other possibilities.  One interesting way to limit the number of
globs that could be resident in the galaxy is through the study of wide angle
binary star systems.  These would be disrupted by close enough passage of
massive mirror globs in the same way they are disrupted by passage of giant
molecular clouds as discussed by Bahcall, Hut and Tremaine (1984)
  There is similar application by Hut and Tremaine (1985)
 to the question of loss of the
sun's Oort cloud, the source of (long-period) comets.  Applying the results of
Bahcall, Hut and Tremaine, we find a time for (tidal) disruption of wide angle 
binary systems by mirror globs somewhat less than the age of the galaxy.  
However, as 
emphasized by Weinberg (1985)
in addition to the uncertainties in our model,
there is the problem of not knowing the initial numbers of wide angle binaries
with enough confidence to say how many have been disrupted. We therefore do
not consider this a reliable test of the model. We note in passing that there is a bound 
(Anderson et al. 1995) on how much nonluminous matter, including mirror matter 
is allowed to exist around the sun.

	Another fruitful area for detecting mirror globs 
would be work  on limiting the number of
halo black holes the galaxy can have, including  observational work such as
searching for X-rays from infalling matter. With $10^6$ globs in the galaxy,
there should be considerable contribution to the X-ray background. In a 
separate, straightforward direction, the above rough model of
the galaxy with a million globs in a 70 kpc sphere would have more than one
glob within 1000 pc of the sun.  One might be able to look at peculiar motions
of stars within such a distance to see whether they are consistent with a
large, but unseen, nearby mass. 

There would be other gravitational effects on galaxies from mirror
globs. Rix and Lake (1993), consider limits on $10^6 M_{\odot}$ objects.
They show that such objects could contribute to dark matter for the Milky
Way but that $10^3 M_{\odot}$ 
is the maximum allowed by relaxational disk heating
in two spiral dwarf galaxies. They point out that this result would not
hold for objects that could be dissipated in the higher rate of collision
for the dwarf case. Their result argues against black compact globs but
not against the other cases. It probably does not rule out case (iii),
since glob-glob scattering might in the dense environment of the dwarf galaxy,
evade the constraint by the liberation from the galaxy of globs and by
evaporation from the globs of some of their mirror hydrogen before collapse.

\vspace{3mm}
\noindent{\Large \bf VI. Conclusion}

\vspace{3mm}

We have analysed the cooling mechanisms for a Jeans
mass glob in the mirror sector as a function of the two free parameters
$\zeta$ and $z_M$ which respectively denote the mass ratio of electrons
in two sectors and the ratio of the redshifts at
which the mirror
globs undergo violent relaxation to that of matter domination. 
We found that for $5\leq \zeta \leq 400$,
cooling is likely to occur by means of the atomic hyperfine structure line.
 This overlaps with the range for $\zeta$ 
(between $10$ to $100$) required to understand
the results of the solar neutrino observation. While we show that the dominant
cooling mechanism in the $ \zeta$  region of interest is too slow to give rise
to glob fragmentation by itself, we cannot rule out with confidence any
of the three end-point cases (at the current age of the universe) of:
(i) puffy globs not greatly condensed, (ii) cluster glob with fragmentation to
low mass bodies or (iii) black glob with collapse to a single black hole.
We noted several possible
ways to test and/or constrain the mirror idea and the three endpoints
 observationally including lensing, wide angle binaries, effect on
relatively local stellar phase space distribution and black glob effects
such as x-ray production.

\vspace{3mm}
\noindent {\bf Acknowledgments}
\vspace{3mm}

One of us (VLT) very much appreciates the hospitality of the University of
Maryland Physics Department, the assistance of 
the Lightner-Sams Foundation in
acquiring the computer on which the numerical work was done and 
several helpful 
conversations with D. Rosenbaum who, inter alia, pointed out the work of
Hut and Tremaine. We are both grateful to D. Armstrong, M. Blecher and 
M. Eckhause for discussion on muonic atoms, to
E. Kolb and A. Bhatia for reading the early draft and especially to D.
Spergel for several stimulating suggestions.

\newpage

\noindent{\Large \bf References}

\vspace{3mm}

\noindent Allison, A. C. and Dalgarno, A. 1969 Ap. J. 158, 423

\vspace{3mm}

\noindent Anderson, J. D. et al. 1995, Ap. J. 448, 885

\vspace{3mm}

\noindent Athanassopoulos, C. et al. 1995, Phys. Rev. Lett. 75, 2650

\vspace{3mm}

\noindent Bahcall, J., Hut, P. and Tremaine, S. 1984, Ap. J. 276, 169

\vspace{3mm}

\noindent Berezhiani, Z. G. and  Mohapatra, R. N. 1995, Phys. Rev. D 52, 6607

\vspace{3mm}

\noindent  Berezhiani, Z. G., Dolgov, A., and Mohapatra, R. N. hep-ph/9511221

\vspace{3mm}

\noindent Binney, J. and Tremaine, S. 1987, {\it Galactic Dynamics},
Princeton University press (Princeton)

\vspace{3mm}

\noindent Bohm-Vitense, E., 1989, {\it Stellar Astrophysics}, volume 1-3,
Cambridge University press (Cambridge)

\vspace{3mm}

\noindent Caldwell, D. and Mohapatra, R. N. 1993, Phys. Rev. D 48, 3259

\vspace{3mm}

\noindent Colombi, S., Dodelson, S., Widrow, L. M. 1996, Ap. J., 458, 1

\vspace{3mm}

\noindent Foot, R. and Volkas, R. 1995, Phys. Rev.  D 52, 6595

\vspace{3mm}

\noindent Green, M., Schwarz, J. and Witten, E. 1986 
{\it Superstring Theories}, Cambridge University press (Cambridge) 

\vspace{3mm}

\noindent Hayashi, K. 1966, Ann. Rev. Astron. and Astrophys., 4, 171

\vspace{3mm}

\noindent Hill, J. 1995, Phys. Rev. Lett., 75, 2655

\vspace{3mm}

\noindent Hodges, H. 1993, Phys. Rev. D47, 456

\vspace{3mm}

\noindent Hughes, V. and Wu, C. S. eds 1975, {\it Muon Physics}, volume 1-3,
Academic Press (NY) 

\vspace{3mm}

\noindent Hut, P. and Tremaine, S. 1985, Astron. J. 90, 1548

\vspace{3mm}

\noindent  Kolb, E. and Turner, M. 1990, {\it The Early Universe }, Addison 
Wesley 

\vspace{3mm}

\noindent  Kolb, E. Seckel, D. and Turner, M. 1985, Nature 314, 415

\vspace{3mm}

\noindent Lang, K. 1980, {\it Astrophysical Formulae}, 
Springer-Verlag (Berlin) 

\vspace{3mm}

\noindent Lepp, S. and Shull, J. M. 1984, Ap. J. 280, 465

\vspace{3mm}

\noindent F. Palla et al. 1983, Ap. J., 271, 632 

\vspace{3mm}

\noindent Peltoniemi, J. and Valle, J. W. F. 1993, Nucl. Phys. B 406, 409

\vspace{3mm}

\noindent Peebles, P. J. E. and Dicke, R. 1968, Ap. J. 154, 891

\vspace{3mm}

\noindent Peebles, P. J. E. 1993, {\it Principles of Physical Cosmology},
Princeton University Press (Princeton)

\vspace{3mm}

\noindent Primack, J. R. et al. 1995, Phys. Rev. Lett., 74, 2160

\vspace{3mm}

\noindent Rix, H-W and Lake, G. 1993, Ap. J. 417, L1

\vspace{3mm}

\noindent Rohlf-Kirsten 1986, {\it Tools in Radio Astronomy}, Springer-Verlag 
(Berlin)

\vspace{3mm}

\noindent Silk, J. 1977, Ap. J. 211, 638 

\vspace{3mm}

\noindent Tegmark, M. et al. 1996, Ap. J. (to be published)

\vspace{3mm}

\noindent Weinberg, M. et al. 1985, Ap. J. 312, 367

\vspace{3mm}

\noindent  Weinberg, S. 1972, {\it Gravitation and Cosmology}, John Wiley
(NY)
\vspace{3mm}

\noindent Winter, K. 1995, CERN Preprint CERN-PPE/95-165 for a recent review
and references

\vspace{4mm}

\newpage
\begin{center}
{\bf Table 1}
\end{center}
\begin{tabular}{|c||c||c||c||c|} \hline
 T & z & $t_U$ & $n_{\gamma}$ & 
$n_{\tilde{B}}~cm^{-3}$ \\ \hline
$T_0=2.7 K$ & 0 & $t_0=4\times 10^{17}$ sec. & 422$cm^{-3}$ 
& $9.4\times 10^{-8}$\\
$T_{R}=0.26 eV$ & 1100 & $1.1\times 10^{13}$ sec. &$ 5.6\times 10^{11}$
 & $1.3\times 10^{2}$\\
$T_{EQ}=1.4$ eV & $5.8\times 10^{3}$ & $7.0\times 10^{11}$ &$ 8.2\times 
10^{13}$ & $1.8\times 10^{4}$\\
$T_{\tilde{R}}$=14($\zeta/30$) eV & $5.8\times 10^{4}(\zeta/30)$& 
$2.9\times 10^{10}(30/\zeta)^2$
& $8.2\times 10^{16}(\zeta/30)^3$ & $1.8\times 10^{7}(\zeta/30)^3$ \\ \hline
\end{tabular}

\noindent {\bf Caption:} 
Early history of mirror matter; here $t_U$ stands for the
age of the universe. 

\begin{center}
{\bf Table 2}

\begin{tabular}{|c||c|} \hline
$M_g$ & $2.9\times 10^{38}$\\
$R_g$ & $3.5\times 10^{18}/ z_M$\\
$\tilde{T}_g$ & $1.8 z_M$ eV\\
$\overline{\rho}_{g}$& $4\times 10^{-19} z^3_M$\\
$\overline{n}_g$ &$ 1.6\times 10^5 z^3_M$\\
$t_g $& $6\times 10^{12} z^{-3/2}_M$ \\ \hline
\end{tabular}

\end{center}

\noindent {\bf Caption:} Glob parameters after violent relaxation. 

\begin{center}
{\bf Table 3}
\end{center}

\begin{tabular}{|c||c||c||c|} \hline

Energy levels & Electronic & vibrational & Rotational \\ \hline

Mirror matter energies & 10$\zeta$ eV & $.2\zeta^{3/2}$ eV & $ 5\times
10^{-3}\zeta^2$ eV \\ \hline
\end{tabular}

\noindent {\bf Caption:} Mirror matter energy levels

\vspace{3mm}

\noindent{\bf Figure Caption:}
 The lines correspond to the three cooling conditions $t_{2,R}=t_0$, 
$t_3=t_0$ and $t_{2,E}=t_0$, where $t_0$ is the age of the universe.
The shaded areas denote the range of the parameters $z_M$ and $\zeta$
where cooling cannot take place before the present age of the universe $t_0$. 

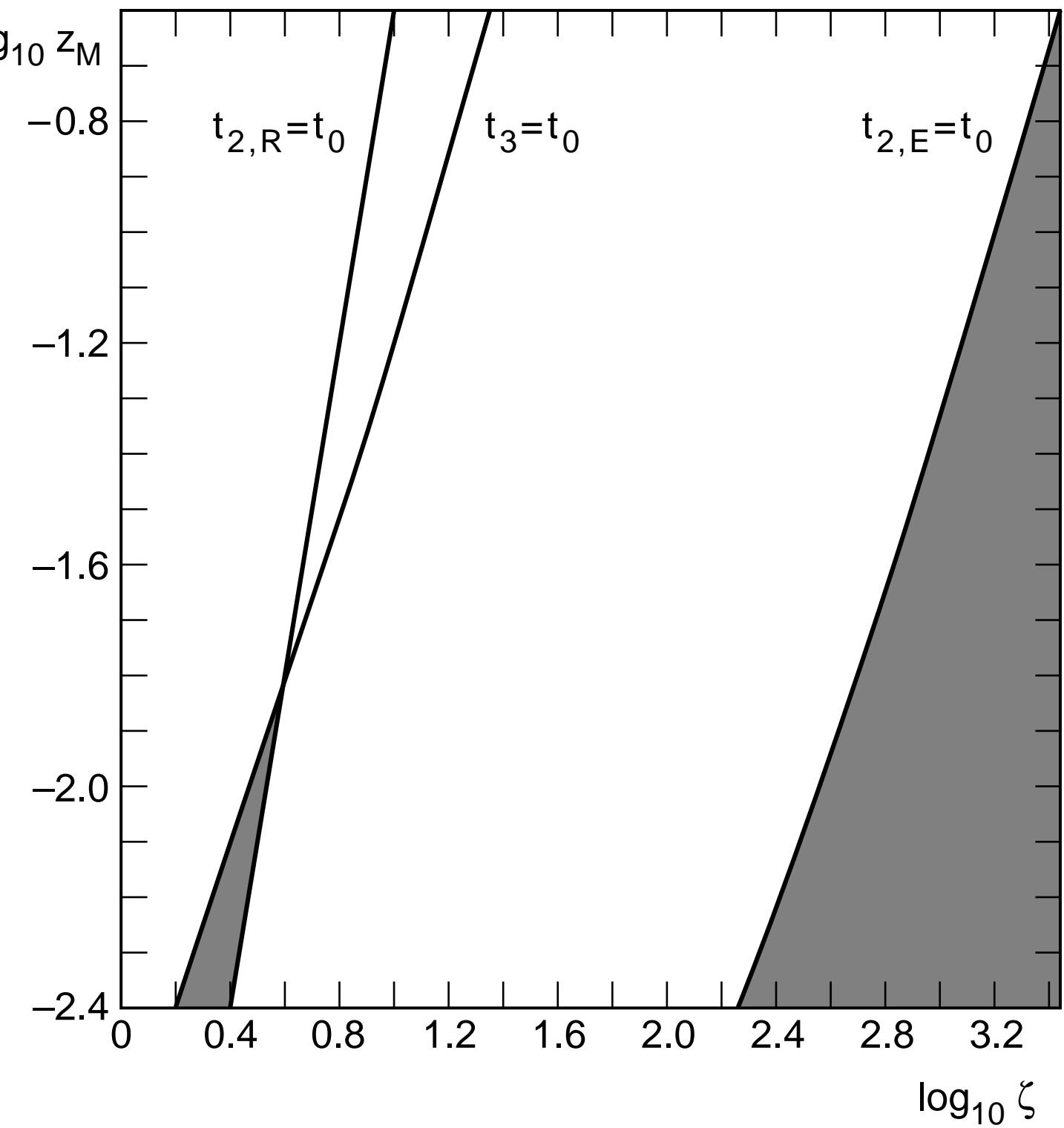
\end{document}